\documentclass[12pt]{article}
\usepackage[dvips]{graphicx}

\begin{document}
\begin{center}
{\bf\Large Two--atomic potential and  distortion of excitation spectrum
of liquid $^4$He by admixture of $^3$He}

\vspace{0.7cm}

{\large\bf D. B. Baranov $^{a}$, A. I. Kirillov $^{b}$,
V. S. Yarunin $^{a,}$
\footnote{Corresponding author. E-mail: yarunin@thsun1.jinr.ru}}

\vspace{0.7cm}

{\it $^a$ Joint Institute for Nuclear Research, Dubna 141980, Moscow Region,
 Russia}\\

{\it $^b$ Moscow Power Engineering Institute, Krasnokazarmennaya 14, E-250,
Moscow 111250, Russia}

\vspace{0.7cm}

PACS: 67.40.-w; 41.20.Cv; 03.65.-w

\vspace{0.7cm}

ABSTRACT
\end{center}

\begin{quote}
The new interatomic potential $g(r)$, providing the tunnelling of atoms,
is suggested for calculating HFB phonon-roton spectrum $E(p)$ of liquid
$^4$He. The tunnelling enables the calculations to reproduce
the  experimental dependence of shifts $E(p)$ on pressure $P$
induced by $^3$He admixture in the constant total density and
volume.
\end{quote}

\vspace{0.5cm}

Distortion of the excitation spectrum of liquid $^4$He by admixture of $^3$He was investigated
experimentally in [1]. The authors of [1] appealed to the theoreticians to
calculate the
excitation spectrums in liquid-helium mixtures in order to explain
their data. The first attempt [2] to explain the $^4$He spectrum shift
theoretically was based on the primary experimental results [3],
that were revised in [1] however.
The recent theory [4] suggested a new version of interatomic potential and
presented the calculation of $^4$He spectrum shift caused by
admixture of $^3$He. Still the result [4] disagreed with the
experiments [1], the explanation of that could be supposed
in the invalidity of Hartree-Fock-Bogoliubov approximation (HFB) in [4].
Now we try to improve the situation using a more realistic potential
than that in [4] in the framework of HFB approximation.
We show that the experimental data can be explained by postulating
that the realistic
potential curve differs from the usually assumed potential curve: the
realistic potential has a local repulsive maximum in the attractive region
of the usual curve. Moreover, the potential depends of the pressure
(see Fig.1).

In this paper, we present an interaction potential $U$, providing the
tunnelling of atoms through the local maximum. This property is inherent for
the atoms in liquids and fills the gap between discrete and
continuous spectrums of ordinary two-atomic potential.

Following [5], we assume that  the polarization effects are not important
in the liquid $^4$He and calculate the interaction potential $U$ between
two $^4$He atoms with electrons in the ground state. We use the formula for
the interaction energy of two atoms with charge distributions densities
$\rho_1 ({\bf r})$ and  $\rho_2 ({\bf r})$
\begin{equation}
\label{one}
U=\int\frac{\rho_1 ({\bf r})\rho_2 ({\bf r' }) d{\bf r}d{\bf r'}}
{| {\bf r}-{\bf r'}|},
\end{equation}
where the charge densities are
\begin{equation}
\begin{array}{l}
\label{two}
\rho_1 ({\bf r})=q\delta ({\bf r}-{\bf r_1})-q F({\bf r}-{\bf r_1}), \\
\phantom{ffff}\\
\rho_2 ({\bf r'})=q\delta ({\bf r'}-{\bf r_2})-q F({\bf r'}-
{\bf r_2}).
\end{array}
\end{equation}
Here $F({\bf r}-{\bf r_1})$ and $F({\bf r'}-{\bf r_2})$ are the electron
probability densities at
the points ${\bf r}$ and ${\bf r'}$, the atom nuclei being at the points
${\bf r_1}$ and ${\bf r_2}$,  and $q=2e$. The interaction between atoms
contains three terms: $U_{nn}$, $U_{nc}$, $U_{cc}$, describing
interactions between two nuclei, nucleus of one atom with the cloud
of the another and between two clouds.

The term $U_{nc}$ has the form
\begin{equation}
\label{nc0}
U_{nc}=-q^2\int\frac{\delta ({\bf r}-{\bf r_1})
 F ({\bf r'}-{\bf r_2}) d{\bf r}d{\bf r'}}
{| {\bf r}-{\bf r'}|}=-
q^2\int\frac{ F ({\bf r'}-{\bf r_2}) d{\bf r'}}
{| {\bf r_1}-{\bf r'}|}=-
q^2\int\frac{ F ({\bf r_0}) d{\bf r_0}}
{|{\bf R}-{\bf r_0}|},
\end{equation}
where ${\bf r_0}={\bf r'}-{\bf r_2}$ and ${\bf R}={\bf r_1}- {\bf r_2}$
is a $3D$ vector between nuclei. The electron probability
density $F(r)$ is determined by  the electron wave function $\Phi$
of the $^4$He atom. Therefore we use the Hylleraas-type wave function with
parameters $\gamma_{1,2}$ (see [6])
$$
\Phi =e^{-s/2a}\left(1+\gamma_1 \frac{u}{a}+
\gamma_2 \frac{t^2}{a^2}\right), \quad s=r_3+r_4, \quad
$$
$$
t=r_3-r_4, \quad u=\left| \bf{r_3}-\bf{r_4}\right|,
$$
where ${\bf r_3}$ and ${\bf r_4}$ are the electron coordinates of atoms
$a$ is the Bohr radius.
In order to obtain the probability density (2) of
the cloud $F$ at the point $r$ one can add the  $w_1$ and $w_2$
probabilities to find one electron at the point $r$, and another at
an arbitrary point
$$
qF(r)=ew_1 (r)+ew_2 (r),
$$
$$
w_1 (r) =\int{\left|\Phi ({\bf r_3 =r, r_4})\right|}^2 d{\bf r_4},
\quad
w_2 (r) =\int{\left|\Phi ({\bf r_3, r_4=r})\right|}^2 d{\bf r_3},
$$

These expressions  show, that the probability density
$F(r)$ can be written as
\begin{equation}
\label{prob}
F(r)=\frac{1}{\pi a^3}e^{-2r/a}(\alpha_4 r^4+\alpha_3 r^3+\alpha_2 r^2+
\alpha_1 r + \alpha_0 + \alpha_{-1} \frac{1}{r})
=\frac{1}{\pi a^3}e^{-2r/a}\psi .
\end{equation}
Thus, the formula {\label{nc0}} for $U_{nc}$ is
\begin{equation}
\label{nc}
U_{nc}=-\frac{q^2}{\pi a^3}\int\frac{e^{-2r_0/a}\psi ({\bf r_0}) d{\bf r_0}}
{|\bf{R}-\bf{r_0}|}.
\end{equation}

The direct computations give
$$
U=\mbox{const}\cdot e^{-2x}\cdot\left(\beta_6{x}^6+
\beta_5{x}^5+\beta_4{x}^4+
\beta_3{x}^3+\right.
$$
\begin{equation}
\label{eq6}
\left. +\beta_2{x}^2+
\beta_1 x+ \beta_0+ \beta_{-1}\frac{1}{x}
\right), \quad x=\frac{R}{a}.
\end{equation}
The parameters $\alpha$ in (\ref{prob}) are the functions of parameters
$\beta$ in (\ref{eq6}). We emphasize that the potential (\ref{eq6})
decreases exponentially at the distances $R\gg a$.

For the sake of simplicity, we limit ourself by considering the
potential of the form
\begin{equation}
g(x)=const\cdot e^{-2x}\left(Ax^6+Bx^5+C+D\frac{1}{x}\right),
\end{equation}
instead of $U(x)$; $g(x)$ has the local maximum and depends on the minimum
number of parameters (see Fig. 1).

\begin{figure}
\centering
\includegraphics[width=14cm, keepaspectratio]{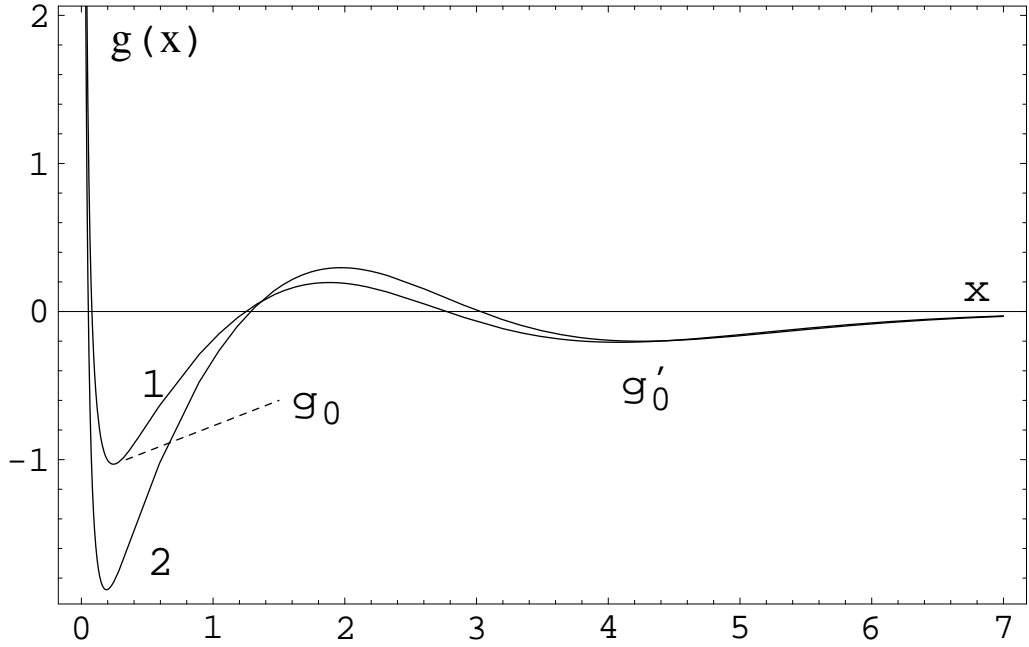}
\caption{Initial (1) and shifted (2) potential  g(x).}
\label{fig11}
\end{figure}

It was shown in [1], that the pressure $P$ increase  in $^4$He
due to the $^3$He admixture (in the same volume $V$ and for the same
total atomic density) leads to the deformation of phonon-roton $^4$He
excitation spectrum $E(p)$. Our purpose is to reproduce
the deformation $\Delta (E)=E(p)|_{P}-E(p)$ of spectrum curves caused by the
pressure $P$. The HFB formula for the spectrum of noncondensate excitation
is
$$
E(p)=\sqrt{\frac{p^4}{4m^2}+\frac{p^2}{m}\rho\tilde g(p)},
$$
where $m$ is the mass of atom, $ \rho$  is the condensate density and
$\tilde g$ is the Fourier transform of $g$. The increase of $P$ leads
to decrease of $\rho$ and to the deformation
of the electron clouds of the atoms, which keep, however, spherically
symmetric. This is evident that the cloud deformation results in the
displacement of the potential minimum $g_0$ "down" and "to the
left" in the $x$, $g(x)$ plane. These curves 1,2 are shown in Fig. 1
and corresponds the values
$$
A_1=-0.5,\, B_1=1.4,\, C_1=-2.5,\, D_1=0.2,
$$
$$
A_2=-0.59,\, B_2=1.8,\, C_2=-3.8,\, D_2=0.2.
$$
The choice of parameters $A, B, C$ and  $D$ is determined by the conditions
for the potential:\\
to be positive (repulsive) for $x\ll 1 $ $(D>0)$,\\
to be negative (attractive) for $x\gg 1$ $(A<0)$,\\
to have local maximum.

The local maximum of $g(x)$ occurs between coordinates of
$g_0$ and $g_0^{'}$. This maximum is introduced in order to describe the
tunnelling of atoms, providing the agreement of the present calculation
with the experimental results for $\Delta (E)$ in [1]. This tunnelling
was not taken into account in [4], and this is the reason for a failure of
[4] to explain [1].

In order to calculate the spectrum, we need the
Fourier transform of $g(x)$
$$
\tilde g(y)=-const\cdot a^3\int\limits_{0}^{\infty}\frac{\sin xy}{xy}g(x)x^2dx=
$$
$$
=-const\cdot a^3
\left[
A\Gamma(8){\left( 4+y^2\right)}^{-4}\sin \left[8\arctan (y/2)\right]
+B\Gamma(7){\left( 4+y^2\right)}^{-7/2}\sin \left[7\arctan (y/2)\right]+
\right.
$$
$$
\left.
+\frac{4C}{{(4+y^2)}^2}+\frac{D}{(4+y^2)}\right], \quad\mbox{where } y=ap.
$$
We let $g_0$ and $g_0^{'}$ denote the minima of the potential $U(x)$
(see Fig. 1), and consider the behavior of $U(x)$.
\begin{figure}
\centering
\includegraphics[width=14cm, keepaspectratio ]{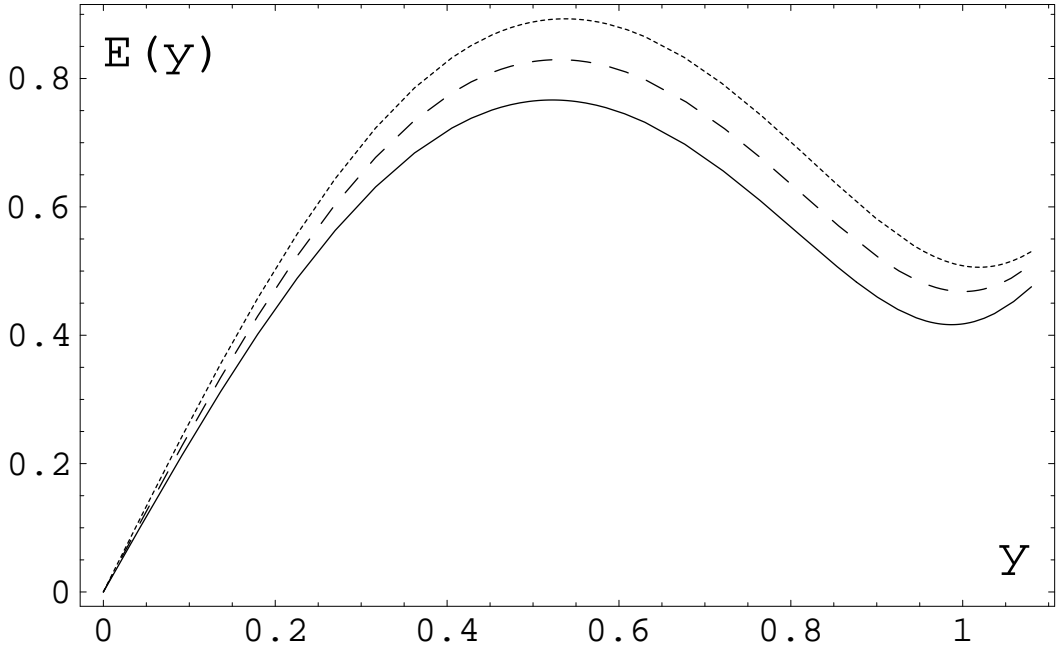}
\caption{Shift of $E(y)$ with the change of
parameters:
$A =-0.5,\, B=1.4,\, C=-2.5,\, D=0.2,\, \tilde{\rho} =0.03$ (solid line);
$A =-0.59,\, B=2.3,\, C=-19,\, D=0.2,\, \tilde{\rho} =0.027$ (dashed line);
$A =-0.61,\, B=4.5,\, C=-30,\, D=0.2,\, \tilde{\rho} =0.025$ (dotted line).}
\label{prep3}
\end{figure}
The curves 1 and 2 in Fig. 1 demonstrate the general situation.

In order to obtain the experimental scale of spectrum shifts we use the
Fourier images of  potentials similar to Fig. 1, but different from
them in scale. Energy excitation spectra for such potentials are shown in
Fig. 2.  The shifted curves (dashed, dotted) in Fig. 2 correspond the
decrease of condensate density $\rho$ and increase of pressure $P$,
relative to the initial state (solid curve). The shifts of the curves in
Fig. 2 qualitatively correspond the deformations of $g(x)$ shown in Fig. 1.

The two types of shifts of initial curve are possible. If the potential
has a local maximum, then there are no intersections of $E(p)$.
We emphasize that if the potential has not a local maximum, then we obtain
the intersections of the shifted spectrum curves as shown in [5].

The local maximum implies the tunnelling of atoms and we conclude that
this property is inherent for quantum liquid.
It means, that in the case of ordinary Lennard--Jones potential there are
discrete and continuous parts of spectrum only. In our case the tunnelling of
atoms is possible in the range of energies intermediate between
continuous and discrete
spectrum. This tunnelling enables us reproduce the experiment sign of
$\Delta (E)$. It describes the atoms, penetrating in liquids through the
energy gap corresponding the regimes of "nearest atoms ordering" (in solids)
and "atomic distant disorder" (in gas).

\vspace{1cm}

D. B. Baranov and V. S. Yarunin are thankful to Russian Foundation
for Basic Research project 00-02-16672 for support.

\phantom{vvvcvccc}
[1] B. F${\rm\stackrel{\circ }{a}}$k, K. Guckelsberger,
M. K$\ddot {\rm o}$rfer, R. Scherm, A. J. Dianoux,  Phys. Rev.

~~~B {\bf 41} 8732 (1990).

[2] W. Hsu, D. Pines, C. H. Aldrich III, Phys. Rev. B
{\bf 32}, 7179 (1985).

[3]  P. A. Hilton, R. Scherm, W. G. Stirling,
J. Low Temp. Phys. {\bf 27}, 851 (1977).

[4] D. Baranov, V. Yarunin, Physica, {\bf A 296}, 337 (2001).

[5] D. B. Baranov, A. I. Kirillov, V. S. Yarunin,
cond-mat/0112440.

[6] H. A. Bethe, E. E. Salpeter, "Quantum mechanics of one-- and
two--electron atoms",

~~~Springer--Verlag, Berlin, 1957.

\end{document}